\documentclass[conference,a4paper]{IEEEtran}
\usepackage[letterpaper, left=0.7in, right=0.7in, bottom=1in, top=0.75in]{geometry}

\usepackage{multicol}
\usepackage{graphicx}
\graphicspath{{Figures/}}
\usepackage{epstopdf}
\usepackage{acronym}
\usepackage{amsmath}
\usepackage{color, colortbl}
\usepackage{subfigure}
\usepackage[english]{babel}
\usepackage{blindtext}
\usepackage{algorithm} 
\usepackage{algorithmic} 
\usepackage{multirow}
\usepackage{color}
\usepackage{subfigure}
\usepackage{balance}
\usepackage{caption}

\begin{document}

\newacro{3GPP}{third generation partnership project}
\newacro{4G}{4{th} generation}
\newacro{5G}{5{th} generation}

\newacro{Adam}{adaptive moment estimation}
\newacro{ADC}{analogue-to-digital converter}
\newacro{AED}{accumulated euclidean distance}
\newacro{AGC}{automatic gain control}
\newacro{AI}{artificial intelligence}
\newacro{AMB}{adaptive multi-band}
\newacro{AMB-SEFDM}{adaptive multi-band SEFDM}
\newacro{AN}{artificial noise}
\newacro{ANN}{artificial neural network}
\newacro{ASE}{amplified spontaneous emission}
\newacro{ASIC}{application specific integrated circuit}
\newacro{AWG}{arbitrary waveform generator}
\newacro{AWGN}{additive white Gaussian noise}
\newacro{A/D}{analog-to-digital}

\newacro{B2B}{back-to-back}
\newacro{BCF}{bandwidth compression factor}
\newacro{BCJR}{Bahl-Cocke-Jelinek-Raviv}
\newacro{BDM}{bit division multiplexing}
\newacro{BED}{block efficient detector}
\newacro{BER}{bit error rate}
\newacro{Block-SEFDM}{block-spectrally efficient frequency division multiplexing}
\newacro{BLER}{block error rate}
\newacro{BPSK}{binary phase shift keying}
\newacro{BS}{base station}
\newacro{BSS}{best solution selector}
\newacro{BU}{butterfly unit}

\newacro{CapEx}{capital expenditure}
\newacro{CA}{carrier aggregation}
\newacro{CBS}{central base station}
\newacro{CC}{component carriers}
\newacro{CCDF}{complementary cumulative distribution function}
\newacro{CCE}{control channel element}
\newacro{CCs}{component carriers}
\newacro{CD}{chromatic dispersion}
\newacro{CDF}{cumulative distribution function}
\newacro{CDI}{channel distortion information}
\newacro{CDMA}{code division multiple access}
\newacro{CI}{constructive interference}
\newacro{CIR}{carrier-to-interference power ratio}
\newacro{CMOS}{complementary metal-oxide-semiconductor}
\newacro{CNN}{convolutional neural network}
\newacro{CoMP}{coordinated multiple point}
\newacro{CO-SEFDM}{coherent optical-SEFDM}
\newacro{CP}{cyclic prefix}
\newacro{CPE}{common phase error}
\newacro{CRVD}{conventional real valued decomposition}
\newacro{CR}{cognitive radio}
\newacro{CRC}{cyclic redundancy check}
\newacro{CS}{central station}
\newacro{CSI}{channel state information}
\newacro{CSIT}{channel state information at transmitter}
\newacro{CSPR}{carrier to signal power ratio}
\newacro{CWT}{continuous wavelet transform}
\newacro{C-RAN}{cloud-radio access networks}

\newacro{DAC}{digital-to-analogue converter}
\newacro{DBP}{digital backward propagation}
\newacro{DC}{direct current}
\newacro{DCGAN}{deep convolutional generative adversarial network}
\newacro{DCI}{downlink control information}
\newacro{DCT}{discrete cosine transform}
\newacro{DDC}{digital down-conversion}
\newacro{DDO-OFDM}{directed detection optical-OFDM}
\newacro{DDO-OFDM}{direct detection optical-OFDM}
\newacro{DDO-SEFDM}{directed detection optical-SEFDM}
\newacro{DFB}{distributed feedback}
\newacro{DFDMA}{distributed FDMA}
\newacro{DFT}{discrete Fourier transform}
\newacro{DFrFT}{discrete fractional Fourier transform}
\newacro{DL}{deep learning}
\newacro{DMA}{direct memory access}
\newacro{DMRS}{demodulation reference signal}
\newacro{DoF}{degree of freedom}
\newacro{DOFDM}{dense orthogonal frequency division multiplexing}
\newacro{DP}{dual polarization}
\newacro{DPC}{dirty paper coding}
\newacro{DSB}{double sideband}
\newacro{DSL}{digital subscriber line}
\newacro{DSP}{digital signal processors}
\newacro{DSSS}{direct sequence spread spectrum}
\newacro{DT}{decision tree}
\newacro{DVB}{digital video broadcast}
\newacro{DWDM}{dense wavelength division multiplexing}
\newacro{DWT}{discrete wavelet transform}
\newacro{D/A}{digital-to-analog}

\newacro{ECC}{error correcting codes}
\newacro{ECL}{external-cavity laser}
\newacro{ECOC}{error-correcting output codes}
\newacro{EDFA}{erbium doped fiber amplifier}
\newacro{EE}{energy efficiency}
\newacro{eMBB}{enhanced mobile broadband}
\newacro{eNB-IoT}{enhanced NB-IoT}
\newacro{EPA}{extended pedestrian A}
\newacro{EVM}{error vector magnitude}

\newacro{Fast-OFDM}{fast-orthogonal frequency division multiplexing}
\newacro{FBMC}{filterbank based multicarrier }
\newacro{FCE}{full channel estimation}
\newacro{FD}{fixed detector}
\newacro{FDD}{frequency division duplexing}
\newacro{FDM}{frequency division multiplexing}
\newacro{FDMA}{frequency division multiple access}
\newacro{FE}{full expansion}
\newacro{FEC}{forward error correction}
\newacro{FEXT}{far-end crosstalk}
\newacro{FF}{flip-flop}
\newacro{FFT}{fast Fourier transform}
\newacro{FFTW}{Fastest Fourier Transform in the West}
\newacro{FHSS}{frequency-hopping spread spectrum}
\newacro{FIFO}{first in first out}
\newacro{F-OFDM}{filtered-orthogonal frequency division multiplexing}
\newacro{FPGA}{field programmable gate array}
\newacro{FrFT}{fractional Fourier transform}
\newacro{FSD}{fixed sphere decoding}
\newacro{FSD-MNSF}{FSD-modified-non-sort-free}
\newacro{FSD-NSF}{FSD-non-sort-free}
\newacro{FSD-SF}{FSD-sort-free}
\newacro{FSK}{frequency shift keying}
\newacro{FTN}{faster than Nyquist}
\newacro{FTTB}{fiber to the building}
\newacro{FTTC}{fiber to the cabinet}
\newacro{FTTdp}{fiber to the distribution point}
\newacro{FTTH}{fiber to the home}

\newacro{GAN}{generative adversarial network}
\newacro{GB}{guard band}
\newacro{GFDM}{generalized frequency division multiplexing}
\newacro{GPU}{graphics processing unit}
\newacro{GSM}{global system for mobile communication}
\newacro{GUI}{graphical user interface}

\newacro{HARQ}{hybrid automatic repeat request}
\newacro{HC-MCM}{high compaction multi-carrier communication}
\newacro{HPA}{high power amplifier}

\newacro{IC}{integrated circuit}
\newacro{ICI}{inter carrier interference}
\newacro{ID}{iterative detection}
\newacro{IDCT}{inverse discrete cosine transform}
\newacro{IDFT}{inverse discrete Fourier transform}
\newacro{IDFrFT}{inverse discrete fractional Fourier transform}
\newacro{ID-FSD}{iterative detection-FSD}
\newacro{ID-SD}{ID-sphere decoding}
\newacro{IF}{intermediate frequency}
\newacro{IFFT}{inverse fast Fourier transform}
\newacro{IFrFT}{inverse fractional Fourier transform}
\newacro{IM}{index modulation}
\newacro{IMD}{intermodulation distortion}
\newacro{IoT}{internet of things}
\newacro{IOTA}{isotropic orthogonal transform algorithm}
\newacro{IP}{intellectual property}
\newacro{IR}{infrared}
\newacro{ISAC}{integrated sensing and communication}
\newacro{ISC}{interference self cancellation}
\newacro{ISI}{inter symbol interference}
\newacro{ISM}{industrial, scientific and medical}
\newacro{ISTA}{iterative shrinkage and thresholding}

\newacro{KNN}{k-nearest neighbours}

\newacro{LDPC}{low density parity check}
\newacro{LFDMA}{localized FDMA}
\newacro{LLR}{log-likelihood ratio}
\newacro{LNA}{low noise amplifier}
\newacro{LO}{local oscillator}
\newacro{LOS}{line-of-sight}
\newacro{LPWAN}{low power wide area network}
\newacro{LS}{least square}
\newacro{LSTM}{long short-term memory}
\newacro{LTE}{long term evolution}
\newacro{LTE-Advanced}{long term evolution-advanced}
\newacro{LUT}{look-up table}

\newacro{MA}{multiple access}
\newacro{MAC}{media access control}
\newacro{MAMB}{mixed adaptive multi-band}
\newacro{MAMB-SEFDM}{mixed adaptive multi-band SEFDM}
\newacro{MASK}{m-ary amplitude shift keying}
\newacro{MB}{multi-band}
\newacro{MB-SEFDM}{multi-band SEFDM}
\newacro{MCM}{multi-carrier modulation}
\newacro{MC-CDMA}{multi-carrier code division multiple access}
\newacro{MCS}{modulation and coding scheme}
\newacro{MF}{matched filter}
\newacro{MIMO}{multiple input multiple output}
\newacro{ML}{maximum likelihood}
\newacro{MLSD}{maximum likelihood sequence detection}
\newacro{MMF}{multi-mode fiber}
\newacro{MMSE}{minimum mean squared error}
\newacro{mMTC}{massive machine-type communication}
\newacro{MNSF}{modified-non-sort-free}
\newacro{MOFDM}{masked-OFDM}
\newacro{MRVD}{modified real valued decomposition}
\newacro{MS}{mobile station}
\newacro{MSE}{mean squared error}
\newacro{MTC}{machine-type communication}
\newacro{MUI}{multi-user interference}
\newacro{MUSA}{multi-user shared access}
\newacro{MU-MIMO}{multi-user multiple-input multiple-output}
\newacro{MZM}{Mach-Zehnder modulator}
\newacro{M2M}{machine to machine}

\newacro{NB-IoT}{narrowband IoT}
\newacro{NB}{naive Bayesian}
\newacro{NDFF}{National Dark Fiber Facility}
\newacro{NEXT}{near-end crosstalk}
\newacro{NFV}{network function virtualization}
\newacro{NG-IoT}{next generation IoT}
\newacro{NLOS}{non-line-of-sight}
\newacro{NLSE}{nonlinear Schrödinger equation}
\newacro{NN}{neural network}
\newacro{NOFDM}{non-orthogonal frequency division multiplexing}
\newacro{NOMA}{non-orthogonal multiple access}
\newacro{NoFDMA}{non-orthogonal frequency division multiple access}
\newacro{NP}{non-polynomial}
\newacro{NR}{new radio}
\newacro{NSF}{non-sort-free}
\newacro{NWDM}{Nyquist wavelength division multiplexing }
\newacro{Nyquist-SEFDM}{Nyquist-spectrally efficient frequency division multiplexing}

\newacro{OBM-OFDM}{orthogonal band multiplexed OFDM}
\newacro{OF}{optical filter}
\newacro{OFDM}{orthogonal frequency division multiplexing}
\newacro{OFDMA}{orthogonal frequency division multiple access}
\newacro{OMA}{orthogonal multiple access}
\newacro{OpEx}{operating expenditure}
\newacro{OPM}{optical performance monitoring}
\newacro{OQAM}{offset-QAM}
\newacro{OSI}{open systems interconnection}
\newacro{OSNR}{optical signal-to-noise ratio}
\newacro{OSSB}{optical single sideband}
\newacro{OTA}{over-the-air}
\newacro{Ov-FDM}{Overlapped FDM}
\newacro{O-SEFDM}{optical-spectrally efficient frequency division multiplexing}
\newacro{O-FOFDM}{optical-fast orthogonal frequency division multiplexing}
\newacro{O-OFDM}{optical-orthogonal frequency division multiplexing}
\newacro{O-CDMA}{optical-code division multiple access}

\newacro{PA}{power amplifier}
\newacro{PAPR}{peak-to-average power ratio}
\newacro{PCA}{principal component analysis}
\newacro{PCE}{partial channel estimation}
\newacro{PD}{photodiode}
\newacro{PDCCH}{physical downlink control channel}
\newacro{PDF}{probability density function}
\newacro{PDP}{power delay profile}
\newacro{PDMA}{polarisation division multiple access}
\newacro{PDM-OFDM}{polarization-division multiplexing-OFDM}
\newacro{PDM-SEFDM}{polarization-division multiplexing-SEFDM}
\newacro{PDSCH}{physical downlink shared channel}
\newacro{PE}{processing element}
\newacro{PED}{partial Euclidean distance}
\newacro{PLA}{physical layer authentication}
\newacro{PLS}{physical layer security}
\newacro{PMD}{polarization mode dispersion}
\newacro{PON}{passive optical network}
\newacro{PPM}{parts per million}
\newacro{PRB}{physical resource block}
\newacro{PSD}{power spectral density}
\newacro{PSK}{pre-shared key}

\newacro{PU}{primary user}
\newacro{PXI}{PCI extensions for instrumentation}
\newacro{P/S}{parallel-to-serial}

\newacro{QAM}{quadrature amplitude modulation}
\newacro{QKD}{quantum key distribution}
\newacro{QoS}{quality of service}
\newacro{QPSK}{quadrature phase-shift keying}
\newacro{QRNG}{quantum random number generation}

\newacro{RAUs}{remote antenna units}
\newacro{RBF}{radial basis function}
\newacro{RBW}{resolution bandwidth}
\newacro{ReLU}{rectified linear units}
\newacro{RF}{radio frequency}
\newacro{RMS}{root mean square}
\newacro{RMSE}{root mean square error}
\newacro{RMSProp}{root mean square propagation}
\newacro{RNTI}{radio network temporary identifier}
\newacro{RoF}{radio-over-fiber}
\newacro{ROM}{read only memory}
\newacro{RRC}{root raised cosine}
\newacro{RSC}{recursive systematic convolutional}
\newacro{RSSI}{received signal strength indicator}
\newacro{RTL}{register transfer level}
\newacro{RVD}{real valued decomposition}

\newacro{SB-SEFDM}{single-band SEFDM}
\newacro{ScIR}{sub-carrier to interference ratio}
\newacro{SCMA}{sparse code multiple access}
\newacro{SC-FDMA}{single carrier frequency division multiple access}
\newacro{SC-SEFDMA}{single carrier spectrally efficient frequency division multiple access}
\newacro{SD}{sphere decoding}
\newacro{SDN}{software-defined network}
\newacro{SDP}{semidefinite programming}
\newacro{SDR}{software-defined radio}
\newacro{SE}{spectral efficiency}
\newacro{SEFDM}{spectrally efficient frequency division multiplexing}
\newacro{SEFDMA}{spectrally efficient frequency division multiple access} 
\newacro{SF}{sort-free}
\newacro{SGD}{stochastic gradient descent}
\newacro{SGDM}{stochastic gradient descent with momentum}
\newacro{SIC}{successive interference cancellation}
\newacro{SiGe}{silicon-germanium}
\newacro{SINR}{signal-to-interference-plus-noise ratio}
\newacro{SIR}{signal-to-interference ratio}
\newacro{SISO}{single-input single-output}
\newacro{SLM}{spatial light modulator}
\newacro{SMF}{single mode fiber}
\newacro{SNR}{signal-to-noise ratio}
\newacro{SP}{shortest-path}
\newacro{SPSC}{symbol per signal class}
\newacro{SPM}{self-phase modulation}
\newacro{SRS}{sounding reference signal}
\newacro{SSB}{single-sideband}
\newacro{SSBI}{signal-signal beat interference}
\newacro{SSFM}{split-step Fourier method}
\newacro{SSMF}{standard single mode fiber}
\newacro{STBC}{space time block coding}
\newacro{STFT}{short time Fourier transform}
\newacro{STO}{symbol timing offset}
\newacro{SU}{secondary user}
\newacro{SVD}{singular value decomposition}
\newacro{SVM}{support vector machine}
\newacro{SVR}{singular value reconstruction}
\newacro{S/P}{serial-to-parallel}

\newacro{TDD}{time division duplexing}
\newacro{TDMA}{time division multiple access }
\newacro{TDM}{time division multiplexing}
\newacro{TFP}{time frequency packing}
\newacro{THP}{Tomlinson-Harashima precoding}
\newacro{TOFDM}{truncated OFDM}
\newacro{TSPSC}{training symbols per signal class}
\newacro{TSVD}{truncated singular value decomposition}
\newacro{TSVD-FSD}{truncated singular value decomposition-fixed sphere decoding}
\newacro{TTI}{transmission time interval}

\newacro{UAV}{unmanned aerial vehicle}
\newacro{UCR}{user compression ratio}
\newacro{UE}{user equipment}
\newacro{UFMC}{universal-filtered multi-carrier}
\newacro{ULA}{uniform linear array}
\newacro{UMTS}{universal mobile telecommunications service}
\newacro{URLLC}{ultra-reliable low-latency communication}
\newacro{USRP}{universal software radio peripheral}

\newacro{VDSL}{very-high-bit-rate digital subscriber line}
\newacro{VDSL2}{very-high-bit-rate digital subscriber line 2}
\newacro{VHDL}{very high speed integrated circuit hardware description language}
\newacro{VLC}{visible light communication}
\newacro{VLSI}{very large scale integration}
\newacro{VOA}{variable optical attenuator}
\newacro{VP}{vector perturbation}
\newacro{VSSB-OFDM}{virtual single-sideband OFDM}
\newacro{V2V}{vehicle-to-vehicle}

\newacro{WAN}{wide area network}
\newacro{WCDMA}{wideband code division multiple access}
\newacro{WDM}{wavelength division multiplexing}
\newacro{WDP}{waveform-defined privacy}
\newacro{WDS}{waveform-defined security}
\newacro{WiFi}{wireless fidelity}
\newacro{WiGig}{Wireless Gigabit Alliance}
\newacro{WiMAX}{Worldwide interoperability for Microwave Access}
\newacro{WLAN}{wireless local area network}
\newacro{WSS}{wavelength selective switch}

\newacro{XPM}{cross-phase modulation}

\newacro{ZF}{zero forcing}
\newacro{ZP}{zero padding}


\title{ Waveform-Defined Privacy:\\A Signal Solution to Protect Wireless Sensing }

\author{\IEEEauthorblockN{ Tongyang Xu}
 \IEEEauthorblockA{Department of Electronic and Electrical Engineering,
University College London, London, UK\\
 Email: {tongyang.xu.11@ucl.ac.uk}}}

\maketitle

\begin{abstract}

Wireless signals are commonly used for communications. Emerging applications are giving new functions to wireless signals, in which wireless sensing is the most attractive one. Channel state information (CSI) is not only the parameter for channel equalization in communications but also the indicator for wireless sensing. However, due to the broadcast nature of wireless signals, eavesdroppers can easily capture legitimate user signals and violate user privacy by measuring CSI. Moreover, the advancement of hardware simplifies illegal eavesdropping since smart devices can track over-the-air signals through walls. Therefore, this work considers a waveform-defined privacy (WDP) solution that can hide CSI phase information and therefore protect user privacy. Besides, the proposed waveform solution achieves better performance due to the use of a unique modulation mechanism. Additionally, by tuning a waveform parameter, the waveform can also enhance communication security.

\end{abstract}


\section{Introduction}

Wireless signals are nowadays everywhere in our daily life. Its ubiquitous deployment not only enables communication-only applications but also boosts wireless sensing applications \cite{Wireless_sensing_Survey_2020} such as human activity detection, health monitoring and presence detection. Recently, an IEEE group is working on a specific standard, termed IEEE 802.11bf \cite{IEEE_802.11bf}, which aims to use existing WiFi signals to realize sensing functions. 

Traditionally, \ac{RSSI} is used for coarse sensing. Recently, \ac{CSI} \cite{WLAN_sensing_JSAC_2017, WLAN_sensing_JIOT_2018, WLAN_sensing_PhaseBeat_ICDCS2017} is applied in fine sensing since CSI has both amplitude and phase information at each sub-carrier. The basic idea is to take advantage of measured \ac{CSI}, in which its variations will be used to sense the change of surrounding environment. Therefore, researchers \cite{WLAN_sensing_ACM_2019} are working on improving sensing accuracy such as improving sensed data quality and using \ac{AI} for better feature extraction. However, user privacy is ignored. Due to the broadcast nature of wireless signals, an eavesdropper can easily capture over-the-air signals in \ac{NLOS} conditions such as behind walls. Since low-cost hardware \ac{SDR} and advanced \ac{AI} are already available on the market, an eavesdropper will therefore easily track human activities or other private user behaviours. The privacy issue will be the main obstacle for ubiquitous deployment of wireless sensing. So far, researchers are mainly focusing on improving sensing accuracy and there are no sufficient efforts on protecting sensing privacy. It is of importance to find a new solution that can protect privacy when signals are omni-directionally broadcasted.

Most of existing research focus on using CSI amplitude \cite{WLAN_sensing_JSAC_2017, WLAN_sensing_JIOT_2018} to learn and track activity patterns and user locations due to its stable characteristics than CSI phase. Due to multipath effects, signal power fading would happen on some sub-carriers. When a person moves with a specific pattern, the signal power fading will change accordingly. Therefore, CSI coefficient amplitude can be used to indicate the movement and location patterns. However, the limited bandwidth of low-frequency wireless signals would merely achieve coarse sensing resolution. For fine resolution sensing, additional information such as CSI coefficient phase should be considered \cite{WLAN_sensing_PhaseBeat_ICDCS2017, WLAN_sensing_PhaseU_Infocom2015}. The better sensing resolution indicates the more sensitive user privacy. Therefore, this work will focus on solutions that can prevent fine resolution sensing.

Wireless sensing highly relies on received signal CSI, therefore a privacy protection solution is to intentionally tune signal waveform characteristics such that an illegal eavesdropper will get incorrect CSI estimation. The tuning aims to change the sign of modulated symbols on some sub-carriers. Therefore, signal spectrum amplitude will not change but the phase of some CSI coefficients will be tuned. Since phase information will determine fine resolution of sensing, its manipulated variations at an eavesdropper will efficiently hide private user behaviours. In addition, by simply tuning sub-carrier spacing without changing sub-carrier bandwidth \cite{Tongyang_CSNDSP_2020}, communication security can be enhanced by the \ac{WDS} framework. Therefore, waveform design can jointly enhance communication security and protect sensing privacy.

The main contributions of this work are as the following.
\begin{itemize}
\item{ Propose a \ac{WDP} solution for wireless sensing by designing a tailored signal waveform.  } 
\item{ The proposed waveform solution ensures reliable communications in terms of \ac{BER}.  } 
\item{ By simply tuning orthogonal waveform features, a non-orthogonal waveform can jointly protect privacy for wireless sensing and enhance communication security.  } 
\end{itemize}

\section{Waveform-Defined Privacy Mechanism} \label{sec:privacy_protection}

\subsection{Privacy Protection} 

This work aims for multicarrier communication scenarios, therefore Fourier transform is required for multicarrier modulation. Assume an $N$-dimension symbol vector $S$ where $N$ is the number of sub-carriers, the modulation at the transmitter is operated in a matrix format as the following 
\begin{equation}\label{eq:tx_matrix_expression}
X=\mathbf{F}S,
\end{equation}
where $\mathbf{F}$ indicates the sub-carrier matrix, which could be replaced by \ac{IFFT} in \ac{OFDM}. The received signal at a receiver side is expressed as
\begin{equation}\label{eq:tx_matrix_Channel_expression}
Y=\mathbf{H}\mathbf{F}S+Z,
\end{equation}
where $\mathbf{H}$ indicates a multipath channel matrix and $Z$ represents \ac{AWGN}. The original signal can be recovered in \eqref{eq:rx_matrix_expression} by multiplying \eqref{eq:tx_matrix_Channel_expression} with $\mathbf{F^*}$
\begin{equation}\label{eq:rx_matrix_expression}
R=\mathbf{F^*}\mathbf{H}\mathbf{F}S+\mathbf{F^*}Z=\mathbf{G}S+W,
\end{equation}
where $R$ indicates the demodulated symbol vector, $\mathbf{G}$ represents a composite matrix, which is a diagonal matrix when \ac{CP} is added in the transmitted signals.

In \ac{WLAN} communications such as 802.11a, the entire frame is termed physical layer conformance procedure (PLCP) protocol data unit (PPDU), which includes legacy preamble and data field. The legacy preamble, consisting of legacy short training field (L-STF), legacy long training field (L-LTF) and legacy signal (L-SIG) field, is used for frequency compensation, phase correction, timing synchronization, channel estimation, \ac{AGC} adjustment, \ac{MCS} notification, etc. The PLCP service data unit (PSDU) in the data field is responsible for carrying data symbols. Typically, the training symbols in legacy preamble is pre-defined and fixed. A receiver will estimate \ac{CSI} based on legacy preambles and recover data symbols using the obtained CSI.

Traditionally, random symbols will be allocated to $S$. However, this work will intentionally manipulate and rearrange each symbol in the vector $S$ such that $S(2)$=-$S(1)$, $S(4)$=-$S(3)$,..., $S(N)$=-$S(N-1)$. Assume the original training symbol vector is $S_t=[S(1), S(2),...,S(N-1), S(N)]$, therefore the \ac{WDP} training symbol vector is obtained as $\bar{S_t}=[S(1), -S(1),...,S(N-1), -S(N-1)]$. The CSI coefficient estimations at a legitimate user and an eavesdropper are computed in \eqref{eq:rx_CSI_eavesdropper} and \eqref{eq:rx_CSI_legitimate} respectively in the following

\begin{subequations}\label{eq:rx_CSI_eavesdropper}
\begin{align}
&h(k)=R_t(k)/S_t(k), \label{eq:rx_CSI_eavesdropper1} \\
&h_a(k)=\sqrt{\Re(h(k))^2+\Im(h(k))^2}, \label{eq:rx_CSI_eavesdropper2} \\
&h_p(k)=tan^{-1}(\frac{\Im(h(k))}{\Re(h(k))}), \label{eq:rx_CSI_eavesdropper3}
\end{align}
\end{subequations}

\begin{subequations}\label{eq:rx_CSI_legitimate}
\begin{align}
&\bar{h}(k)=R_t(k)/\bar{S}_t(k), \label{eq:rx_CSI_legitimate1} \\
&\bar{h}_a(k)=\sqrt{\Re(\bar{h}(k))^2+\Im(\bar{h}(k))^2}, \label{eq:rx_CSI_legitimate2}\\
&\bar{h}_p(k)=tan^{-1}(\frac{\Im(\bar{h}(k))}{\Re(\bar{h}(k))}), \label{eq:rx_CSI_legitimate3}
\end{align}
\end{subequations}
where $k=1,2,...,N$, $R_t(k)$ indicates the $k^{th}$ demodulated training symbol, $S_t(k)$ is the $k^{th}$ original training symbol, $\bar{S}_t(k)$ is the $k^{th}$ WDP training symbol, $h(k)$ and $\bar{h}(k)$ are the $k^{th}$ estimated complex CSI coefficient at an eavesdropper and a legitimate user, respectively. $h_a(k)$ and $\bar{h}_a(k)$ indicate the $k^{th}$ CSI coefficient amplitude at the eavesdropper and the legitimate user, respectively. $h_p(k)$ and $\bar{h}_p(k)$ indicate CSI coefficient phase at the eavesdropper and the legitimate user, respectively. 

When $k$ is an odd number belonging to $[1,3,5,...,N-1]$, both the legitimate user and the eavesdropper will estimate the same CSI coefficient. Assume $R_t(k)=c+di$, $S_t(k)=a+bi$ and $\bar{S}_t(k)=a+bi$, where $|a|$=$|b|$=1 for QPSK modulations, the complete computations for one CSI coefficient at an odd-index $k$ are the following
\begin{subequations}\label{eq:rx_CSI_specific_odd_index}
\begin{align}
&h(k)=\bar{h}(k)=\frac{(ac+bd)+(ad-bc)i}{a^2+b^2}, \label{eq:rx_CSI_specific_odd_index1} \\
&h_a(k)=\bar{h}_a(k), \label{eq:rx_CSI_specific_odd_index2} \\
&h_p(k)=\bar{h}_p(k). \label{eq:rx_CSI_specific_odd_index3}
\end{align}
\end{subequations}

When $k$ is an even number belonging to $[2,4,6,...,N]$, the eavesdropper will estimate a different CSI coefficient. Assume $R_t(k)=u+vi$, $S_t(k)=p+qi$ and $\bar{S}_t(k)=-(a+bi)$, where $|a|$=$|b|$=$|p|$=$|q|$=1 for QPSK modulations, the complete computations for one CSI coefficient at an even-index $k$ are the following
\begin{subequations}\label{eq:rx_CSI_specific_even_index}
\begin{align}
&h(k)=\frac{(up+vq)+(vp-uq)i}{p^2+q^2}, \label{eq:rx_CSI_specific_even_index1}\\
&\bar{h}(k)=\frac{-(ua+vb)-(va-ub)i}{a^2+b^2}, \label{eq:rx_CSI_specific_even_index2}\\
&h_a(k)=\bar{h}_a(k), \label{eq:rx_CSI_specific_even_index3}\\
&h_p(k)\neq\bar{h}_p(k). \label{eq:rx_CSI_specific_even_index4}
\end{align}
\end{subequations}

Therefore, based on the expressions in \eqref{eq:rx_CSI_specific_odd_index} and \eqref{eq:rx_CSI_specific_even_index}, It is apparent that the eavesdropper and the legitimate user will estimate the same CSI coefficient amplitude and phase at odd-index sub-carriers while different CSI coefficient phase will be obtained at even-index sub-carriers.

\subsection{Communication Reliability} 

Wireless sensing privacy is protected by modulating opposite-sign symbols on adjacent even-index and odd-index sub-carriers. Its communication reliability is studied in this section. Considering the mathematical expression in \eqref{eq:rx_matrix_expression}, the computations for the first demodulated symbol and the second demodulated symbol are given in \eqref{eq:ICI_analysis_tx_cancellation_first_symbol} and \eqref{eq:ICI_analysis_tx_cancellation_second_symbol}, respectively.
\begin{subequations} \label{eq:ICI_analysis_tx_cancellation_first_second_symbol} 
\begin{align}
&R'(1)=\sum_{k=1{\&} k{\in}odd}^{N-1}S(k)[\mathbf{G}(1,k)-\mathbf{G}(1,k+1)]+W(1), \label{eq:ICI_analysis_tx_cancellation_first_symbol} \\
&R'(2)=\sum_{k=1{\&} k{\in}odd}^{N-1}S(k)[\mathbf{G}(2,k)-\mathbf{G}(2,k+1)]+W(2). \label{eq:ICI_analysis_tx_cancellation_second_symbol}
\end{align}
\end{subequations}

A general expression for each demodulated symbol is given as
\begin{equation}\label{eq:ICI_analysis_tx_cancellation_general}
R'(\varphi)=\sum_{k=1{\&} k{\in}odd}^{N-1}S(k)[\mathbf{G}(\varphi,k)-\mathbf{G}(\varphi,k+1)]+W(\varphi).
\end{equation}

Therefore, the general expression for the new composite matrix is defined as

\begin{equation}\label{eq:ICI_analysis_tx_cancellation_first_symbol_ICI}
\mathbf{G}'(\varphi,k)=\mathbf{G}(\varphi,k)-\mathbf{G}(\varphi,k+1),
\end{equation}
where $\varphi=1,2,3,...,N$. Since symbols at even-index sub-carriers are opposite-sign copies of the symbols at odd-index sub-carriers, therefore effective demodulated symbols $R'(\varphi)$ should be indexed as $\varphi=1,3,5,...,N-1$. The signal power for odd-index effective symbols can be further increased via additional receiver side operations as the following.

\begin{equation} \label{eq:ICI_analysis_tx_rx_cancellation_first_symbol}
\begin{split}
  R''(1)&=R'(1)-R'(2) \\
  &=\sum_{k=1{\&} k{\in}odd}^{N-1}S(k)[\mathbf{G}(1,k)-\mathbf{G}(1,k+1)-\\
  &\;\ \ \  \mathbf{G}(2,k)+\mathbf{G}(2,k+1)]+W'(1),
\end{split}
\end{equation}
where $W'(1)=W(1)-W(2)$. Since $\mathbf{G}$ is a Toeplitz matrix, therefore $\mathbf{G}(1,k)$=$\mathbf{G}(2,k+1)$ and a new expression is given as

\begin{equation} \label{eq:ICI_analysis_tx_rx_cancellation_first_symbol_new_express}
\begin{split}
  R''(1)&=R'(1)-R'(2) \\
  &=\sum_{k=1{\&} k{\in}odd}^{N-1}S(k)[2\mathbf{G}(1,k)-\mathbf{G}(1,k+1)-\\
  &\;\ \ \  \mathbf{G}(2,k)]+W'(1).
\end{split}
\end{equation}

Then a general expression for $R''$ is given as

\begin{equation} \label{eq:ICI_analysis_tx_rx_cancellation_general_symbol}
\begin{split}
  \underbrace{R''(\xi)}_{\xi{\in}odd}&=R'(\xi)-R'(\xi+1) \\
  &=\sum_{k=1{\&} k{\in}odd}^{N-1}S(k)[2\mathbf{G}(\xi,k)-\mathbf{G}(\xi,k+1)-\\
  &\;\ \ \  \mathbf{G}(\xi+1,k)]+W'(\xi).
\end{split}
\end{equation}

Thus a new composite component is expressed as 

\begin{equation}\label{eq:ICI_analysis_rx_tx_cancellation_general_symbol_ICI}
\mathbf{G}''(\xi,k)=2\mathbf{G}(\xi,k)-\mathbf{G}(\xi,k+1)-\mathbf{G}(\xi+1,k) .
\end{equation}

\section{Communication Security} \label{sec:communication_security}

Traditionally, data can be secured via advanced encryption methods such as \ac{QRNG} and \ac{QKD} \cite{quantum_QKD_PRL_1991, quantum_UK_GOV_2020_white_paper, quantum_satellite_China_Science_2017}. In this case, even though data is captured by eavesdroppers, they cannot easily understand the hidden messages. However, encryption cannot prevent adversarial attacks such as using \ac{AI} to intentionally manipulate over-the-air signals such that a legitimate user cannot recover received signals using a correct decryption key. Fortunately, waveform design can deal with the security challenge as well. By tuning the traditional OFDM waveform to a non-orthogonal waveform \cite{TongyangTVT2017}, a \ac{WDS} framework \cite{Tongyang_CSNDSP_2020} can work efficiently together with the proposed \ac{WDP} scheme. In this case, communication security and sensing privacy can be achieved simultaneously via waveform design. 

As defined in \eqref{eq:tx_matrix_expression}, $\mathbf{F}$ is an $N\times{N}$ sub-carrier matrix with elements equal to $e^{\frac{j2{\pi}{nk}}{N}}$. It is clear that adjacent sub-carriers (e.g. sub-carrier index $n_1$, $n_2$) are orthogonally packed and an eavesdropper will easily decode the signal. One defence idea is to introduce a bandwidth compression factor $0<\alpha<1$ in $e^{\frac{j2{\pi}{nk}\alpha}{N}}$. The continuous and fractional features of $\alpha$ will complicate signal decoding since an eavesdropper will not be able to detect the exact value of $\alpha$ \cite{tongyang_VTC2020_DL_classification}. Therefore, the computation of \eqref{eq:rx_matrix_expression} at an eavesdropper will fail since the exact $\mathbf{F^*}$ is not available. With a mismatched demodulation matrix $\mathbf{F^*}$, signals cannot be properly recovered by an eavesdropper.

\section{System Modelling and Simulation Results}

To simplify system modelling, this work considers single-antenna multicarrier communication scenarios covering both \ac{OFDM} and the non-orthogonal signal waveforms, in which the number of data sub-carriers is 64 and the total number of time samples is 128. As explained in Section \ref{sec:privacy_protection}, half of sub-carriers are used to carry opposite-sign copies of original symbols, therefore the effective number of data sub-carriers is 32.

To test the CSI hidden capability of using the proposed signal opposite-sign modulation scheme, a randomly modelled multi-path channel is defined as the following
\begin{eqnarray}\label{eq:static_channel}
{{h(t)}=0.8655\delta(t)+0.255e^{\frac{-j\pi}{2}}\delta(t-3T_s)}\nonumber
\\{-0.4312e^{\frac{j\pi}{2}}\delta(t-5T_s)},
\end{eqnarray}
where the channel has three paths and the maximum time delay is 5$T_s$ where $T_s$ indicates the time interval of one time sample. The channel model is static and its \ac{PDP} is configured intentionally such that a signal experiences multi-path frequency selective channel distortions. The channel model could be any other configurations.

\begin{figure}[t]
\begin{center}
\includegraphics[scale=0.65]{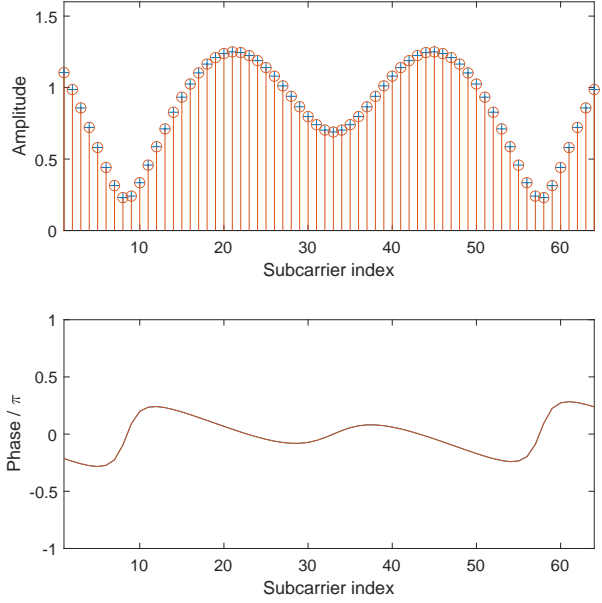}
\end{center}
\caption{CSI amplitude and phase characteristics at a legitimate user based on the OFDM signal format.}
\label{Fig:WDP_traditional_amplitude_phase}
\end{figure}

\begin{figure}[t]
\begin{center}
\includegraphics[scale=0.65]{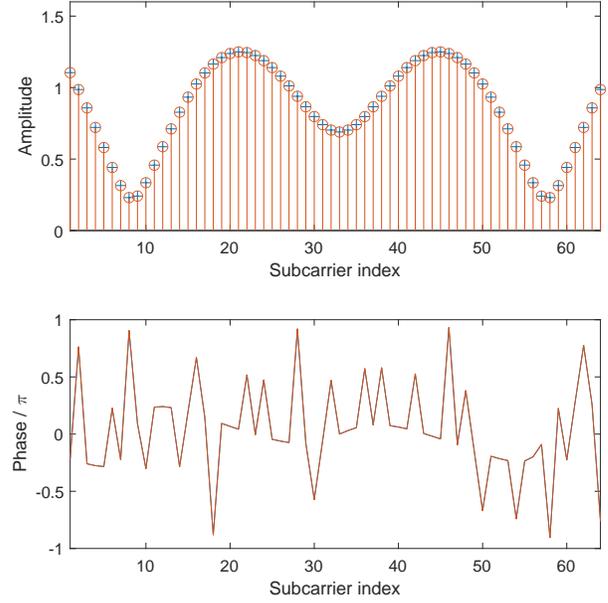}
\end{center}
\caption{CSI amplitude and phase characteristics at an eavesdropper based on the OFDM signal format.}
\label{Fig:WDP_proposed_amplitude_phase}
\end{figure}

The spectral amplitude and phase characteristics at the legitimate user are illustrated in Fig. \ref{Fig:WDP_traditional_amplitude_phase}, in which each impulse represents a \ac{CSI} coefficient amplitude at a specific sub-carrier. It is apparent that the CSI amplitude illustration is non-flat and the signal experiences multi-path effects. In addition, its phase coefficient at each sub-carrier location is different and can be extracted for sensing purposes. On the other hand, the CSI coefficient amplitude and phase at an eavesdropper are illustrated in Fig. \ref{Fig:WDP_proposed_amplitude_phase}. It is obvious that the eavesdropper can extract the same CSI amplitude information as the legitimate user. However, it will obtain different CSI phase information. This is due to the manipulated symbols at even-index sub-carriers. To ensure a convincing CSI estimation at the eavesdropper side, an eavesdropper can rely on practical spectrum analyzers to check their estimated CSI. However, the eavesdropper cannot easily recognize the difference since the estimated CSI amplitude shows similar spectral shape to the one obtained at the legitimate user. Therefore, the eavesdropper will not notice the CSI coefficient phase variations and will fail to get valid sensing information.

\begin{figure}[t]
\begin{center}
\includegraphics[scale=0.56]{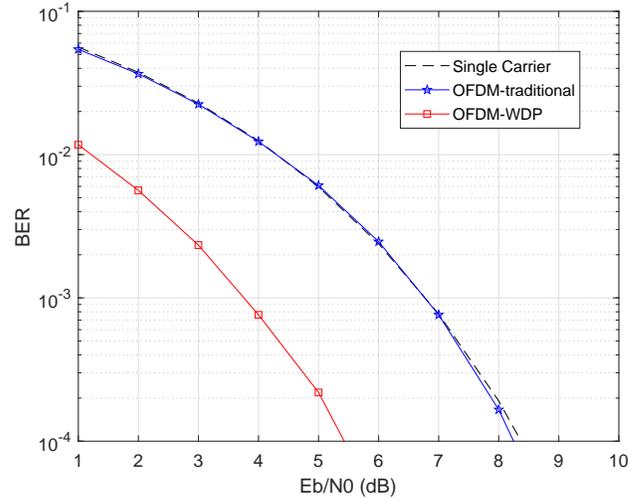}
\end{center}
\caption{BER performance for the WDP signal in AWGN channel.}
\label{Fig:WDP_BER}
\end{figure}

Communication reliability is first studied in an \ac{AWGN} channel. Fig. \ref{Fig:WDP_BER} shows BER performance for traditional OFDM and the proposed OFDM-WDP signals. Due to the enhanced signal quality at each odd-index sub-carrier according to the calculation in \eqref{eq:ICI_analysis_tx_rx_cancellation_general_symbol}, the proposed OFDM-WDP outperforms the traditional OFDM by approximately 3 dB at BER=$10^{-4}$. It is noted that the proposed waveform outperforms single-carrier waveform since the proposed WDP signal employs a repetition coding similar modulation pattern where the even-index symbols are opposite-sign copies of odd-index symbols. 

\begin{figure}[t]
\begin{center}
\includegraphics[scale=0.56]{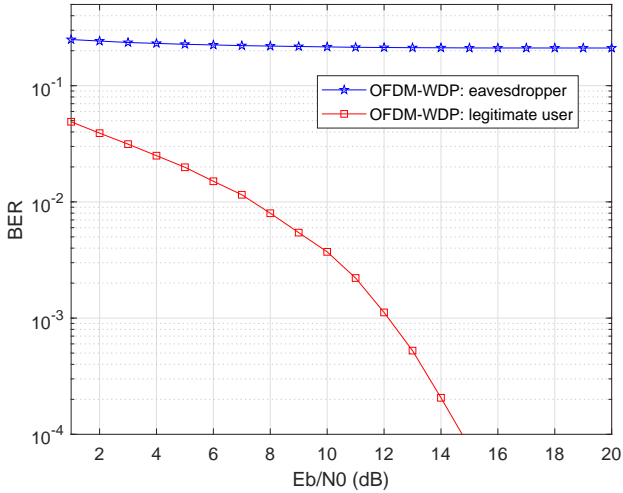}
\end{center}
\caption{BER performance for WDP signals at the eavesdropper and the legitimate user after the frequency selective channel distortions.}
\label{Fig:WDP_BER_channel}
\end{figure}

Considering the multi-path channel model in \eqref{eq:static_channel}, the BER performance for each signal waveform under the frequency selective channel distortion is compared in Fig. \ref{Fig:WDP_BER_channel}. To show stable and convincing BER performance, ideal CSI is assumed to be known at equalization. It is obvious that the proposed WDP signal is robust at the legitimate user while the performance at the eavesdropper is significantly degraded. This is due to the CSI phase errors at even-index sub-carriers at the eavesdropper in Fig. \ref{Fig:WDP_proposed_amplitude_phase}.

\begin{figure}[t]
\begin{center}
\includegraphics[scale=0.56]{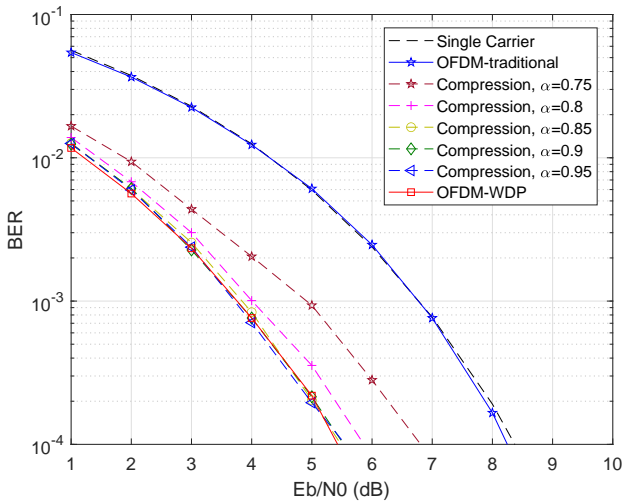}
\end{center}
\caption{BER performance for different values of $\alpha$ at a legitimate user.}
\label{Fig:WDP_BER_WDS}
\end{figure}

\begin{figure}[t]
\begin{center}
\includegraphics[scale=0.56]{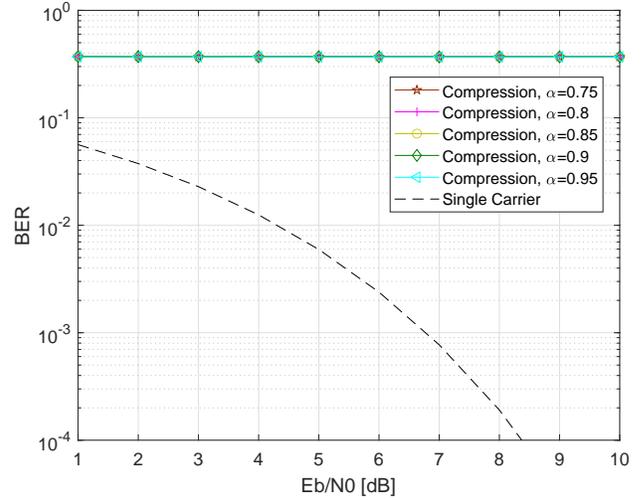}
\end{center}
\caption{BER performance at an eavesdropper when exact values of $\alpha$ are not known.}
\label{Fig:WDP_BER_WDS_fake}
\end{figure}

Communication security is studied in Fig. \ref{Fig:WDP_BER_WDS} and Fig. \ref{Fig:WDP_BER_WDS_fake}. The values of bandwidth compression factor $\alpha$ are defined as 0.95, 0.9, 0.85, 0.8, 0.75. It is shown in Fig. \ref{Fig:WDP_BER_WDS} that the signals at $\alpha$=(0.95, 0.9, 0.85) achieve the same performance as the proposed OFDM-WDP signal. With further reduction of $\alpha$, performance loss starts to appear. On the other hand, Fig. \ref{Fig:WDP_BER_WDS_fake} reveals that when an eavesdropper has no information of the exact value of $\alpha$, signals will not be correctly decoded and error floors appear.

\section{Conclusion}

In wireless sensing, applications are typically relying on signal power variations rather than the exact data carried by signals. However, signal power merely shows signal strength without showing its phase information. Therefore, pure amplitude information is suitable for coarse sensing applications. For fine sensing applications, additional phase information is required, which is normally obtained by measuring signal \ac{CSI}. This work proposed a waveform-defined privacy (WDP) modulation framework such that legitimate user privacy is protected by hiding CSI phase information. Results revealed that by modulating opposite-sign symbols on adjacent even-index sub-carriers and odd-index sub-carriers, an eavesdropper can extract correct CSI coefficient amplitude information but fail to obtain correct CSI coefficient phase information. In terms of communication reliability, results revealed that the repetition coding similar symbol modulation mechanism helps to achieve better BER performance than traditional orthogonal signals. In addition to sensing privacy and communication reliability, results also verified that communication security is ensured by simply tuning waveform parameters such that an eavesdropper cannot properly decode signals.

\bibliographystyle{IEEEtran}
\bibliography{VTC2021_Ref}

\end{document}